\documentclass[twoside,10pt,a4paper]{newFNLstyle}
\usepackage{graphics}
\usepackage{cite}

\begin{document}

\volnumpagesyear{0}{0}{000--000}{2001}
\dates{received date}{revised date}{accepted date}

\title{INVITATION TO THE \lq\lq SPOOKY" QUANTUM PHASE-LOCKING EFFECT AND ITS LINK
TO $1/F$ FLUCTUATIONS}

\authorsone{MICHEL PLANAT \thanks{Use footnotes only to indicate if permanent and present addresses are different. Funding information should go in the Acknowledgement section.}}
\affiliationone{Laboratoire de Physique et M\'{e}trologie des
Oscillateurs du CNRS,\\ 32 Avenue de l'observatoire, 25044
Besan\c{c}on Cedex, France\\planat@lpmo.edu}

%\authorstwo{FOURTH AUTHOR}
%\affiliationtwo{Full affiliation}
%\mailingtwo{and mailing address}

% Add author/affliation/mailing sets as required

\maketitle

\markboth{Invitation to the \lq\lq spooky" quantum phase-locking
effect and its link to $1/f$ fluctuations}{}

\pagestyle{myheadings}
% Comment this out to remove the running heads

% Keywords have to before the abstract I'm afraid.

\begin{abstract}
An overview of the concept of phase-locking at the non linear,
geometric and quantum level is attempted, in relation to finite
resolution measurements in a communication receiver and its $1/f$
noise. Sine functions, automorphic functions and cyclotomic
arithmetic are respectively used as the relevant trigonometric
tools. The common point of the three topics is found to be the
Mangoldt function of prime number theory as the generator of low
frequency noise in the coupling coefficient, the scattering
coefficient and in quantum critical statistical states. Huyghens
coupled pendulums, the Adler equation, the Arnold map, continued
fraction expansions, discrete M\"obius transformations, Ford
circles, coherent and squeezed phase states, Ramanujan sums, the
Riemann zeta function and Bost and Connes KMS states are some but
a few concepts which are used synchronously in the paper.
\end{abstract}

\section{Introduction}

The interleaving of frequencies and phases of electronic
oscillators interacting in non linear circuits follows
arithmetical rules. Continued fraction expansions, prime number
decompositions and related number theoretical concepts were
successfully used to account for the experimental effects in
mixers and phase-locked loops\cite{FNL01,APL02}. We also made use
of these tools within the field of quantum optics emphasizing the
hidden connection between phase-locking and cyclotomy\cite{PLA03}.
As a matter of fact the understanding of phase effects in devices
is so much entangled that it will reveal useful to check several
clues relying on differential and discrete non linear equations,
Fourier analysis, hyperbolic geometry and quantum mechanics. But
over all the fundamental issue is the modelling of finite
resolution measurements and how it puts a constraint on the
performance, how it generates the fluctuations in oscillating
circuits.

In Sect. \ref{Classical} we report on the early history of
phase-locking about classical observations of coupled mechanical
pendulums and electronic oscillators. Non linear continuous and
discrete generic models are introduced emphasizing the case of
homodyne detection in a phase sensitive communication receiver. It
is shown that low frequency noise happens due to low pass
filtering and the finite resolution of frequency counts. A
phenomenological model relating the coupling coefficient to prime
numbers is developed. In Sec \ref{Hyperbolic} the justification of
the model is given by using the hyperbolic geometry of the
half-plane applied to the low pass filtering. In Sec \ref{Quantum}
the phase itself is taken to be discrete. Discrete quantum optics
is related to quantum phase-locking and the arithmetic of $1/f$
noise.

\section{Classical Phase-Locking: from Huyghens to the Prime Numbers}
\label{Classical}

{\it Being obliged to stay in my room for several days and also
occupied in making observations on my two newly made clocks, I
have noticed a remarkable effect which no one could have ever
thought of. It is that these two clocks hanging next to one
another separated by one or two feet keep an agreement so exact
that the pendulums invariably oscillate together without
variation. After admiring this for a while, I finally figured out
that it occurs through a kind of sympathy: mixing up the swings of
the pendulums, I have found that within a half hour always return
to consonance and remain so constantly afterwards as long as I let
them go. I then separated them, hanging one at the end of the room
and the other fifteen feet away, and noticed that in a day there
was five seconds difference between them. Consequently, their
earlier agreement must in my opinion have been caused by an
imperceptible agitation of the air produced by the motion of the
pendulums.}

 The citation is taken from \cite{LaserPhysics74}. The authors remind a
later letter by Huyghens that the coupling mechanism was in fact a
small vibration transmitted through the wall, and not movement of
air:

 {\it Lord Rayleigh (1907) made similar observations about two
driven tuning forks coupled by vibrations transmitted through the
table on which both forks sat... Locking in triode circuits was
explained by Van der Pol (1927) who included in the equation for
the triode oscillator an external electromotive force as given in
\begin{equation}
\frac{d^2v}{dt^2}-\frac{d}{dt}(gv-\beta' v^3)+\omega^2
v=\omega_0^2 V_0 \sin \omega_0 t,
\end{equation}
where $g$ is the linear net gain (i.e. the gain in excess of
losses, $\beta'$ the saturation coefficient, and $\omega$ is the
resonance frequency in the absence of dissipation or gain. He
showed that when an external electromotive force is included, of
frequency $\omega_0$, and tuned close to the oscillator frequency
$\omega$, the oscillator suddenly jumped to the external
frequency. It is important to note that the beat note between the
two frequencies vanishes not because the two frequencies vanish,
not because the triode stops oscillating, but because it
oscillates at the external frequency.

We can show the locking effect by utilizing the slowly varying
amplitude approach, including a slowly varying phase $\Phi$ and
oscillation at the external frequency $\omega_0$ and amplitude $V$
\begin{equation}
\frac{d\Phi}{dt}+K \sin \Phi= \omega-\omega_0=\omega_{LF},
\label{Adler}
\end{equation}
where we use $\omega_{LF}$ for the detuning term and $K=\omega_0
V_0/V$ for the locking coefficient}\cite{LaserPhysics74}.

The regime just described is the so-called injection locking
regime, also found in injection-locked lasers. The equation
(\ref{Adler}) is the so-called Adler's equation of
electronics\cite{Adler46}.

One way to synthesize (\ref{Adler}) is thanks to the phase-locked
loop of a communication receiver. The receiver is designed to
compare the information carrying external oscillator (RF) to a
local oscillator (LO) of about the same high frequency through a
non linear mixing element. For narrow band demodulation one uses a
discriminator of which the role is first to differentiate the
signal, that is to convert frequency modulation (FM) to amplitude
modulation (AM) and second to detect its low frequency envelope:
this is called baseband filtering. For more general FM
demodulation one uses a low pass filter instead of the
discriminator to remove the high frequency signals generated after
the mixer. In the closed loop operation a voltage controlled LO
(or VCO) is used to track the frequency of the RF. Phase
modulation is frequently used for digital signals because low bit
error rates can be obtained despite poor signal to noise ratio in
comparison to frequency modulation\cite{Klapper72}.

Let us consider a type of receiver which consists in a mixer, in
the form of a balanced Schottky diode bridge and a low pass
filter. If $f_0$ and $f$ are the frequencies of the RF and the LO,
and $\theta(t)$ and $\psi(t)$ their respective phases, the set
mixer and filter essentially behaves as a phase detector of
sensitivity $u_0$ (in Volts/rad.), that is the instantaneous
voltage at the output is the sine of the phase difference at the
inputs
\begin{equation}
u(t)=u_0\sin(\theta(t)-\psi(t)).
\end{equation}
The non linear dynamics of the set-up in the closed loop
configuration is well described by introducing the phase
difference $\Phi(t)=sin(\theta(t)-\psi(t))$. Using
$\dot{\theta}=\omega_0$ and $\dot{\psi(t)}=\omega+Au(t)$, with
$\omega_0=2\pi f_0$, $\omega=2\pi f$ and A (in rad. Hz/Volt) as
the sensitivity of the $VCO$, one recovers Adler's equation
(\ref{Adler}) with the open loop gain $K=u_0A$.

Equation (\ref{Adler}) is integrable but its solution looks
complex\cite{DosSantos98}. If the frequency shift $\omega_{LF}$
does not exceed the open loop gain $K$, the average frequency
$\langle \dot{\Phi} \rangle$ vanishes after a finite time and
reaches the stable steady state
$\Phi(\infty)=2l\pi+\sin^{-1}(\omega_{LF}/K)$, $l$ integer. In
this phase-tracking range of with $2K$ the $RF$ and the $LO$
oscillators are also frequency-locked. Outside the mode-locking
zone there is a sech shape beat signal of frequency
\begin{equation}
\tilde{\omega}_{LF}=\langle \dot{\Phi}(t)
\rangle=(\omega_{LF}^2-K^2)^{1/2}. \label{lowfrequency}
\end{equation}
\begin{figure}[htbp]
\centering{\resizebox{7cm}{!} {\includegraphics{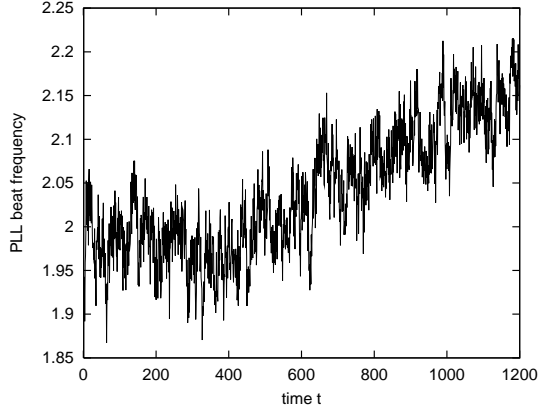}}}
\caption{Fluctuating counts of the beat frequency (in $Hz$) close
to the phase-locked zone. The inputs are quartz oscillators at
$10$ MHz. The power spectrum has a pure $1/f$ dependance}
\end{figure}
The sech shape signal and the non linear dependance on parameters
$\omega_{LF}$ and $K$ are actually found in
experiments\cite{DosSantos98,APL02}. In addition the frequency
$\omega_{LF}$ is fluctuating (see Fig. 1). It can be characterized
by the Allan variance $\sigma^2(\tau)$ which is the mean squared
value of the relative frequency deviation between adjacent samples
in the time series, averaged over an integration time $\tau$.
Close to the phase-locked zone the Allan deviation is
\begin{equation}
\sigma(\tau)=\frac{\sigma_0 K}{\tilde{\omega}_{LF}},
\label{1surfnoise}
\end{equation}
where $\sigma_0$ is a residual frequency deviation depending of
the quality of input oscillators and that of the phase detector.
Allan deviation is found independent of $\tau$ which is a
signature of a $1/f$ frequency noise of power spectral density
$S(f)=\sigma/(2 \ln 2 f)$. One way to predict the dependence
(\ref{1surfnoise}) is to use differentiation of
(\ref{lowfrequency}) with respect to the frequency shift
$\tilde{\omega_{LF}}$ so that
\begin{equation}
\delta\tilde{\omega}_{LF}=\delta\omega_{LF}
(1+K^2/\tilde{\omega}_{LF}^2)^{1/2}.\label{deltaLF}
\end{equation}
Relation (\ref{deltaLF}) is defined outside the mode-locked zone
$|\omega_{LF}|>K$; close to it, if the effective beat note
$\tilde{\omega}_{LF}\le K$, the square root term is about
$K/\tilde{\omega}_{LF}$. If one identifies $\delta
\omega_{LF}/\tilde{\omega}_{LF}$ with a pedestal Allan deviation
$\sigma_0$ and $\delta \tilde{\omega}_{LF}/\tilde{\omega}_{LF}$
with a magnified Allan deviation $\sigma$ one explains the
experimental result (\ref{1surfnoise}). One can conclude that,
either the PLL set-up behaves as a microscope of an underlying
flicker floor $\sigma_0$, or the $1/f$ noise is some dynamical
property of the PLL. In the past we looked at a possible low
dimensional structure of the time series and found a stable
embedding dimension lower or equal to $4$\cite{IEEE00}. But at
that time the dynamical model of $1/f$ noise still remained
elusive.

Adler's model presupposes a fundamental interaction
$\omega_{LF}=|\omega_0-\omega(t)|$ in the mixing of the two input
oscillators. But the practical operation of the phase detector
involves harmonic interactions of the form $\omega_{LF}=|p
\omega_0 -q \omega(t)| \le \omega_c=2\pi f_c$, where $p$ and $q$
are integers and $f_c$ is the cut-off frequency of the low pass
filter. This can be rewritten by introducing the frequency ratios
$\nu=\frac{\omega (t)}{\omega_0}$ and
$\mu=\frac{\omega_{LF}}{\omega_0}$ as $\mu=q|\nu-\frac{p}{q}|$.
This form suggests that the aim of the receiver is to select such
couples $(p,q)$ which realize a \lq\lq good" approximation of the
\lq\lq real" number $\nu$. There is a mathematical concept which
precisely does that: the diophantine approximator. It selects such
couples $p_i$ and $q_i$, coprime to each other, i.e. with greatest
common divisor $(p_i,q_i)=1$ from the continued fraction expansion
of $\nu$
\begin{equation}
\nu=[a_0;a_1,a_2,\cdots
a_i]=a_0+1/(a_1+1/(a_2+1/\cdots+1/(a_i\cdots)))\simeq\frac{p_i}{q_i}.
\label{confrac}
\end{equation}
The diophantine approximation satisfies
\begin{equation}
|\nu-\frac{p_i}{q_i}|\le\frac{1}{a_{i+1}q_i^2}. \label{dioph}
\end{equation}
The fraction $\frac{p_i}{q_i}$ is a so-called convergent and the
$a_i'$'s are called partial quotients. The approximation is
truncated at the index $i$ just before the partial quotient
$a_{i+1}$. It should be observed that diophantine approximations
are different from decimal approximations $\frac{c_i}{d_i}$ for
which one gets $|\nu-\frac{c_i}{d_i}|\le \frac{1}{d_i}$. It was
shown\cite{FNL01} using the filtering condition that $a_{i+1}$
identifies with a very simple expression
\begin{equation}
a_{i+1}=\left[ \frac{f_0}{f_c q_i}\right], \label{filter}
\end{equation}
where $[~ ]$ denotes the integer part. For example if one chooses
$f_0=10$ MHz and $f_c=300$ kHz, the fundamental basin
$\frac{p_i}{q_i}=\frac{1}{1}$ will be truncated if $a_{i+1}\ge 33$
and the basin $\frac{p_i}{q_i}=\frac{3}{5}$ will be truncated if
$a_{i+1}\ge 6$. The resulting full spectrum is a superposition of
V-shape basins of which the edges are located at
\begin{eqnarray}
&\nu_1=\{a_0;,a_1,a_2,\cdots,a_i,a_{i+1}\},\nonumber\\
&\nu_2=\{a_0;a_1,a_2,\cdots,a_{i-1},1,a_{i+1}\}, \label{edge}
\end{eqnarray}
where the partial expansion before $a_{i+1}$ corresponds to the
two possible continued fractions of the rational number
$\frac{p_i}{q_i}$. The basin of number
$\nu=\frac{3}{5}=\{0;1,1,2\}$ extends to
$\nu_1=\{0;1,1,2,33\}=\frac{19}{32}\simeq 0.594 $,
$\nu_2=\{0;1,1,1,1,33\}=\frac{31}{34}\simeq 0.618$. For a
reference oscillator with $f_0=10$ MHz this corresponds to a
frequency bandwidth $(0.618-0.594).10^7$ MHZ=240 kHz.

With these arithmetical rules in mind one can now tackle the
difficult task to account for phase-locking of the whole set of
harmonics. The differential equation for the phase shift
$\dot{\Phi}(t;q_i,p_i)$ at the harmonic $(p_i,q_i)$ corresponding
to the beat frequency
\begin{equation}
\omega_{LF}=|p_i\omega_0-q_i\omega(t)|, \label{beat}
\end{equation}
can be obtained as
\begin{eqnarray}
&&\dot{\Phi}(t;q_i,p_i)+q_i\mbox{ }H(P)\mbox{
}\sum_{r_i,s_i}K(r_i,s_i) \nonumber\\ &&\times
\sin(\frac{s_i}{q_i}\mbox{ }\Phi(t;q_i,p_i)-\frac{\omega_0
t}{q_i}(q_i r_i-p_i s_i)+\Phi_0(r_i,s_i))
=\omega_{\rm{LF}}(p_i,q_i). \label{Adlerpq}
\end{eqnarray}
The notation $K(r_i,s_i)$ means the effective gain at the harmonic
$r_i/s_i$, $\Phi_0(r_i,s_i)$ is the reference angle and $H(P)$,
where the operator $P=\frac{d}{dt}$, is the open loop transfer
function. Solving (\ref{Adlerpq}) is a difficult task. Let us
observe that the RF signal at frequency $\omega_0$ acts as a
periodic perturbation of the Adler's model of the PLL. If one
neglects harmonic interactions, (\ref{Adlerpq}) may be simplified
to the standard Arnold map model
\begin{equation}
\Phi_{n+1}=\Phi_n+2\pi\Omega-c~ \sin \Phi_n, \label{Arnold}
\end{equation}
where $\Omega=\frac{\omega}{\omega_0}$ is the bare frequency ratio
and $c=\frac{K}{\omega_0}$. Such a nonlinear map is studied by
introducing the winding number $\nu=\lim _{n \rightarrow \infty }
(\Phi_n-\Phi_0)/(2 \pi n)$. The limit exists everywhere as long as
$c<1$, the curve $\nu$ versus $\omega$ is a devil's staircase with
steps attached to rational values $\Omega=\frac{p_i}{q_i}$ and
width increasing with the coupling coefficient $c$. The
phase-locking zones may overlap if $c>1$ leading to chaos from
quasi-periodicity\cite{Cvit92}.

The Arnold map is also a relevant model of a short Josephson
junction shunted by a strong resistance $R$ and driven by a
periodic current of frequency $\omega_0$ and amplitude $I_0$.
Steps are found at the driving voltages
$V_r=RI_0=r(\hbar\omega_0/2e)$, $r$ a rational number. Fundamental
resonances $r=n$, $n$ integer, have been used to achieve a voltage
standard of relative uncertainty $10^{-7}$.

To appreciate the impact of harmonics on the coupling coefficient
one may observe that each harmonic of denominator $q_i$ creates
the same noise contribution $\delta \omega_{LF}=q_i \delta
\omega(t)$. They are $\phi(q_i)$ of them, where $\phi(q_i)$ is the
Euler totient function, that is the number of integers less or
equal to $q_i$ and prime to it; the average coupling coefficient
is thus expected to be $1/\phi(q_i)$. In \cite{APL02} a more
refined model is developed based on the properties of prime
numbers. It is based on defining a coupling coefficient as $c^*=c
\Lambda(n;q_i,p_i) $ with $\Lambda(n;q_i,p_i)$ a generalized
Mangoldt function. Mangoldt function is defined as
\begin{equation}
\Lambda(n)=\Lambda(n;1,1)= \left\{\begin{array}{ll}
\ln b &~~\mbox{if}~ n=b^k,~b~\mbox{a~prime},\\
 0 &~~ \mbox{otherwise}.\\
\end{array}\right.
\label{Mangoldt}
\end{equation}
The generalized Mangoldt function attached to the resonance
$\frac{p_i}{q_i}$ adds the restriction that $n$ should also be
congruent to $p_i \rm{mod}(q_i)$. The important result of that
analysis is to exhibit a fluctuating average coefficient as
follows
\begin{equation}
c_{\rm{av}}^*/c=\frac{1}{t} \sum_{n=1}^{t}
\Lambda(n;q_i,p_i)=\frac{1}{\phi(q_i)}+\epsilon(t)
,\label{flucgain}
\end{equation}
with $\epsilon(t)=O(t^{-1/2}\ln^2(t))$ which is known to be a good
estimate as long as $q_i<\sqrt t$\cite{FNL01}. The average
coupling coefficient shows the expected dependance on $q_i$. In
addition there is an arithmetical noise $\epsilon(t)$ with a low
frequency dependance of the power spectrum reminding $1/f$ noise.
Although that stage of the theory is not the last word of the
story, it is quite satisfactory that this approach, based on
phase-locking of the full set of harmonics, is accounting for the
main aspects of $1/f$ noise found in experiments.

\section{Hyperbolic Phase-Locking}
\label{Hyperbolic}

{\it The concept of an automorphic function is natural
generalization of that of a periodic function. Furthermore, an
automorphic form is a generalisation of the exponential function
$e(z)=\exp(2i\pi z)$.}

 The citation is from Iwaniec's book \cite{Iwaniec02}.

 As shown in the previous section the homodyne detector behaves as
 a diophantine approximator of the frequency ratio $\nu$ of input
 oscillators. The approach is very satisfactory in explaining
 frequency-locking effects in the open loop, but the recourse to
 non linear differential equations is necessary for the case of
 the closed loop. Now we attempt to develop a pure arithmetical
 frame to account for the phase-locking effects as well. This is
 done by normalizing the beat frequency with respect to the low
 pass cut-off frequency as $y=\frac{\omega_{LF}}{\omega_c}$
 instead of the reference frequency. This is suggested by the
 geometry of continued fraction expansions which resorts to new
 mathematical concepts such as Ford circles, the hyperbolic
 half-plane, M\"obius transformations, hyperbolic Laplace
 equation...

 The hyperbolic half-plane is defined as the set of complex
 numbers $z$ such that $\Im z\ge 0$. It is a rich mathematical
 object first studied by the mathematician Henri Poincar\'e at the
 end of nineteenth century. The geometry of $\mathcal{H}$ is
 related to continued fraction expansions  and it is also a
 natural frame to study the prime numbers. It will be used also for
  studying phase-locking effects. We
 put\footnote{The imaginary number $i$ such that $i^2=-1$ should not be confused with
 the index $i$ in integers $p_i$, $q_i$ and related integers.}.
\begin{equation}
z=\nu+iy,~~\nu=\frac{\omega}{\omega_0},~0<y=\frac{\omega_{LF}}{\omega_c}<1.\label{complexplane}
\end{equation}
It is clear that $y>0$ is a condition which is imposed by counting
measurements. The second condition $y<1$ results from the low pass
filtering and will reveal important in our introduction of Ford
circles below. Let us start with the continued fraction expansion
 (\ref{confrac}) which is rewritten as
$\{a_0;a_1,\cdots a_i\}=a_0+\frac{1}{\{a_1;a_2\cdots
a_i\}}=\frac{p_i}{q_i}$.

 By induction $p_0=a_0$, $q_0=1$,
$\{a_0,a_1\}=a_0+\frac{1}{a_1}=\frac{a_0a_1+1}{a_1}=\frac{p_1}{q_1}$,
so that
\begin{eqnarray}
&\left[\begin{array}{cc} a_0 & 1\\ 1 & 0
\end{array}\right] \left[\begin{array}{cc} a_1 & 1\\ 1 & 0
\end{array}\right]=\left[\begin{array}{cc} p_1&p_0\\
q_1&q_0
\end{array}\right],\nonumber\\
&\left[\begin{array}{cc} a_0 & 1\\ 1 & 0 \end{array}\right]
\left[\begin{array}{cc} a_1 & 1\\ 1 & 0 \end{array}\right]\cdots
\left[\begin{array}{cc} a_i & 1\\ 1 & 0 \end{array}\right]
=\left[\begin{array}{cc} p_i&p_{i-1}\\ q_i&q_{i-1}
\end{array}\right]. \label{matrices}
\end{eqnarray}
Taking determinants and using a propery of continued fractions one
gets $p_iq_{i-1}-p_{i-1}q_i=(-1)^{i-1}$. This suggests to
associate to the convergents $\frac{p_i}{q_i}$ the group of
M\"obius transformations
\begin{equation}
z \rightarrow z'=\gamma(z)=\frac{p_i z+p_{i-1}}{q_i z+q_{i-1}},
~~p_iq_{i-1}-p_{i-1}q_i=1. \label{Mobius}
\end{equation}
A M\"obius transform map a point $z\in \mathcal{H}$, with
$\mathcal{H}$ the upper half-plane $\Im z >0$ into a point
$z'\in\mathcal{H}$. M\"obius transforms form a group $\Gamma$
called the modular group. It is a discontinuous group that is used
to tesselate $\mathcal{H}$ by copies of a fundamental domain
$\mathcal{F}$ under the group action. The fundamental domain of
$\Gamma$ (or modular surface) is defined as $\mathcal{F}=\{z \in
\mathcal{H}:~|z|\ge 1,~|\nu|\le \frac{1}{2}\}$, and the family of
domains $\{\gamma(\mathcal{F}),\gamma \in \Gamma\}$ induces a
tesselation of $\mathcal{H}$ \cite{Iwaniec02}.

We are also interested by the Ford circles\cite{ICNF03}: they are
defined by the images of the filtering line $z=\nu+i$ under all
M\"obius transformations (\ref{Mobius}). Rewriting (\ref{Mobius})
as $q_iz'-p_i=-1/(q_iz+q_{i-1})$ and inserting $z=\nu+i$ in that
expression one immediately gets
\begin{equation}
|z'-(\frac{p_i}{q_i} + \frac{i}{2q_i^2})|=\frac{1}{2q_i^2}.
\label{Ford}
\end{equation}
which are circles of radius $\frac{1}{2q_i^2}$ centered at
$\frac{p_i}{q_i} + \frac{i}{2q_i^2}$. To each $\frac{p_i}{q_i}$ a
Ford circle in the upper-half plane can be attached, which is
tangent to the real axis at $\nu=\frac{p_i}{q_i}$. Ford circles
never intersect: they are tangent to each other if and only if
they belong to fractions which are adjacent in the Farey sequence
$\frac{0}{1}<\cdots\frac{p_1}{q_1}<\frac{p_1+p_2}{q_1+q_2}<\frac{q_1}{q_2}\cdots<\frac{1}{1}$.
Ford circles can be considered a complex plane view of continued
fractions.

Ford circles are also related to the wavefronts of an hyperbolic
noise model: the Laplace equation for $\mathcal{H}$. To see this
one first observe that $\mathcal{H}$ carries a non-Euclidean
metric $dz=(d\nu^2+dy^2)^{1/2}$. This is easily shown by using
$\Im \gamma(z)=\frac{y}{|q_i z+q_{i-1}|^2}$ and
$\frac{d\gamma(z)}{dz}=\frac{1}{(q_iz+q_{i-1})^2}$, since
$|\frac{d\gamma(z)}{dz}|=\frac{\Im \gamma(z)}{y}$ under all group
actions.

The invariance of the metric is inherited by the non-Euclidean
Laplacian
\begin{equation}
\Delta=y^2 (\frac{\partial^2}{\partial \nu^2}+
\frac{\partial^2}{\partial y^2}). \label{Laplacian}
\end{equation}
We will be concerned by eigenfunctions $\Psi_s(z)$ of
(\ref{Laplacian}) as our dynamical model of phase-locking. The
complex parameter $s$ has been introduced as a label for the
eigenvalues. The relevant wave equation is
\begin{equation}
(\Delta+\lambda)\Psi_s(z)=0. \label{Laplace}
\end{equation}
The solutions of (\ref{Laplace}) are called automorphic forms.

They are several ways of finding eigenfunctions $\Psi_s(z)$ for a
given eigenvalue $\lambda$. We will first consider elementary
power law solutions, then we will show that there is a method to
generate a lot of eigenvalues out of a fixed $\Psi_s(z)$ by
shifting to $\Psi_s(\gamma(z))$, and still more by averaging over
selected $\gamma(z)$ in $\Gamma$.

The simplest solutions of (\ref{Laplace}) are horizontal waves of
eigenvalue $\lambda=s(s-1)$ in the form of power laws
\begin{equation}
\Psi_s(z)=\left\{\begin{array}{ll}
y^s\\
 y^{1-s}\\
\end{array}.
\right .
 \label{powerlaw}
\end{equation}
Any other wave which satisfies (\ref{Laplace}) can be searched by
shifting $z$ to $\gamma(z)$ so that $y$ goes to $\Im(\gamma(z))$
with the result
\begin{equation}
\Psi_s(z)=(\Im \gamma(z))^s=\frac{y^s}{|q_iz+q_{i-1}|^{2s}}.
\label{solutions}
\end{equation}
The wavefronts for that solution are obtained by assigning to the
quantity $y/(q_i z +q_{i-1})$ some constant value $c$. As a result
one gets the equation
$(x+q_{i-1}/q_i)^2+(y-1/(2q_i^2c^2))^2=1/(4c^4q_i^4)$ of circles
tangent to the real axis at $x=-q_{i-1}/q_i$ and of radius
$1/(2c^2q_i^2)$. At $c=1$ such a circle is obtained from the Ford
circle of index $i$ by an horizontal slip from $\nu=p_i/q_i$ to
$q_{i-1}/q_i$. It contracts ($c<1$) or expands ($c>1$) remaining
tangent to the real axis.

The more general solution of (\ref{Laplace}) is obtained by
summing solutions (\ref{solutions}) over $\gamma$ transformations
in $\Gamma$ taken over the right coset $\Gamma_{\infty}\ \setminus
\Gamma$ where $\Gamma_{\infty}$ is the subgroup of translations $z
\rightarrow z+n$. One gets\cite{Knauf99}
\begin{equation}
\Psi_s(z)=y^s(1+\sum_i \frac{1}{|q_iz+q_{i-1}|^{2s}}),
\label{general}
\end{equation}
where the summation index $i$ means that the summation should be
applied to all Farey fractions $\frac{p_i}{q_i}$.

The whole solution can be decomposed into three contributions, the
horizontal wave $y^s$ is incident onto the filtering line, the
reflected wave $S(s)y^{1-s}$ is scattered with a scattering
coefficient $S(s)$ of modulus $1$, whereas the remaining part
$T_s(y,\nu)$ is a complex superposition of waves depending on $y$
and the harmonics of $\exp(2i\pi\nu)$, but goes to zero away from
the real line. More precisely one obtains $S(s)=A(s)Z(s)$,
$Z(s)=\frac{\zeta(2s-1)}{\zeta(2s)}$,
$A(s)=\frac{\Gamma(1/2)\Gamma(s-1/2)}{\Gamma(s)}$, $\Gamma(s)$ is
the Gamma function and $\zeta(s)=\sum_{n=1}^{\infty}\frac{1}{n^s}$
is the Riemann zeta function.
\begin{figure}[htbp]
\centering{\resizebox{8cm}{!}{\includegraphics{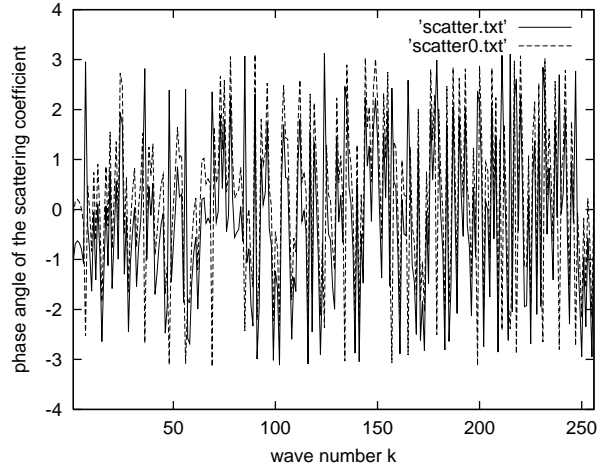}}}
\caption{The phase angle $\kappa(k)$ for the scattering of noise
waves on the modular surface. Plain lines: Exact phase factor.
Dotted lines: Approximation based on the quotient $A(k)$ of two
Riemann zeta functions at $2ik$ and $2ik+1$.}
\end{figure}
There is much to say about the singularities of the Riemann zeta
function. It is analytic only on that part of complex plane where
$\Re s>1$. But in the scattering coefficient it appears in the
extended Riemann zeta function
$\xi(s)=\pi^{-s/2}\Gamma(s/2)\zeta(s)$ which is analytic over the
whole complex plane. The scattering coefficient equals the
quotient $S(s)=\xi(2s-1)/\xi(2s)$. One knows from the theory of
zeta function that $s=\frac{1}{2}$ is an axis of symmetry for
$\zeta(s)$, and according to Riemann hypothesis all non trivial
zeros lie on that axis, with a very random distribution. An
interesting case is thus to compute the scattering coefficient
$S(s)$ along the critical line $s=\frac{1}{2}+ik$ in which case
the superposition $T_s(y,\nu)$ also vanishes. As a result
\begin{equation}
S(k)\propto \exp(2i\kappa(k))~~\mbox{with}~\kappa'(k)=\frac{d\ln
Z(s) }{ds}~~\mbox{at}~s=\frac{1}{2}+ik. \label{scatter}
\end{equation}
In that expression we removed the smooth part $A(k)$ in the
scattering coefficient so that its modulus is no longer $1$ but
its phase variability is still well described by the exponential
factor above (see Fig. 2). The aim of this replacement is to get
an explicit link of the counting function $\kappa'(k)$ to the
Mangoldt function $\Lambda(n)$ already accounted in
(\ref{Mangoldt}) as an effective coupling coefficient in the
Arnold map (\ref{Arnold}). It is known that the logarithmic
derivative of $\zeta(s)$ is
$-\frac{\zeta'(s)}{\zeta(s)}=\sum_{n\ge 1}\frac{\Lambda(n)}{n^s}$.
It is also easy to check that with
$B(s)=\frac{\zeta(s)}{\zeta(s+1)}$, one gets
$-\frac{B'(s)}{B(s)}=\sum_{n\ge 1}\frac{b(n)}{n^s}$ with
$b(n)=\frac{\Lambda(n)\phi(n)}{n}$ the modified Mangoldt function.
Here $Z(s)=B(2s-1)$.

It is now clear that the scattering of waves in the hyperbolic
plane has much to share with the phase-locking model formulated in
Sect. \ref{Classical}. One argument in favor of that new frame is
that the average modified Mangoldt function satisfies
$B(t)=\frac{1}{t}\sum_{n\ge 1}b(n)=1+\epsilon_B(t)$, where
$\epsilon_B(t)$ is an error term of power spectral density equal
to $1/f^{2G}$ with $G\simeq 0.618$ the Golden ratio\cite{ICNF03}.

\section{Quantum Phase-Locking}
\label{Quantum}

{\it Apparently Dirac was the first to attempt a definition of a
phase operator by means of an operator amplitude and phase
decomposition. As we have discussed, with a complex $c$-number
$a=Re^{i\Phi}$ one obtains the phase via $e^{i\Phi}=a/R$.
Similarly, he sought to decompose the annihilation operator $a$
into amplitude and phase components... After a brief calculation
we obtain a relation indicating that the number operator $N$ and
phase operator $\Phi$ are canonically conjugate
\begin{equation}
[N,\Phi]=1.
\end{equation}
The equation immediately leads to a number-phase uncertainty
relation which is often seen
\begin{equation}
\delta N ~\delta \Phi \ge 1/2.
\end{equation}
However, all of the previous development founders upon closer
examination.}

 This is taken from \cite{Lynch95}, a comprehensive review of the quantum
 phase problem. See also\cite{Schleich93}.

 To approach the phase-locking problem within quantum mechanics one
 can start from the theory of the harmonic oscillator. The natural
 objects are the Fock states (the photon occupation states)
 $|n\rangle$ who live in an infinite dimensional Hilbert space.
 They are orthogonal to each other: $\langle n|m\rangle
 =\delta_{mn}$, where $\delta_{mn}$ is the Dirac symbol. The states form a
 complete set: $\sum_{n=0}^{\infty}|n\rangle \langle n|=1$.

 The annihilation operator removes one photon from the electromagnetic field
\begin{equation}
a|n\rangle=\sqrt{n}|n-1\rangle, n=1,2,\cdots
\end{equation}
Similarly the creation operator $a^{\dag}$ adds one photon:
$a^{\dag}|n\rangle=\sqrt{n+1}|n+1\rangle, n=0,1,\cdots$ There is
the commutation relation $[a,a^{\dag}]=1$. The operator $N=a
a^{\dag}$ has the meaning of the particle number operator and
satisfies the eigenvalue equation $N|n\rangle=n|n\rangle$.

Eigenvalues of the annihilation operator are the so-called
coherent states $|\alpha\rangle$\cite{Dodonov03}
\begin{equation}
a|\alpha\rangle=\alpha|\alpha\rangle,~~\mbox{width}~|\alpha\rangle=e^{-|\alpha|^2/2}
\sum_{n=0}^{\infty} \frac{\alpha^n |n\rangle }    {(n!)^{1/2}} ,
\end{equation}
where the eigenvalue $\alpha$ is a complex parameter which
quantifies the intensity of the field. A single mode laser
operated well above threshold generates a coherent state
excitation. The electric field variation of a coherent state
approaches that of a classical wave of stable amplitude and fixed
phase. In the coordinate representation it is a minimal
uncertainty state.

States of well defined phase escaping the inconsistencies of
Dirac's formulation were build by Susskind and
Glogower\cite{Susskind64}. They correspond to the eigenvalues of
the exponential operator
\begin{equation}
E=e^{i\Phi}=(N+1)^{-1/2}a=\sum_{n=0}^{\infty}|n\rangle\langle
n+1|.
\end{equation}
Using the Hermitian conjugate operator $E^{\dag}=e^{-i\Psi}$, one
gets $E E^{\dag}=1$, $E^{\dag}E=1-|0\rangle\langle0|$, i.e. the
unitarity of $E$ is spoiled by the vacuum-state projector
$|0\rangle\langle0|$. The Susskind-Glogower phase states satisfy
the eigenvalue equation $E|\Psi\rangle=e^{i\psi}|\Psi\rangle$;
they are given as
\begin{equation}
|\Psi\rangle=\sum_{n=0}^{\infty}e^{in\psi}|n\rangle.
\label{Susskind}
\end{equation}
Like the coherent states the phase states are non orthogonal and
they form an overcomplete basis which solves the identity
operator: $\frac{1}{2\pi}\int_{-\pi}^{\pi}d\psi|E\rangle\langle
E|=1$. The operator $\cos \Phi=\frac{1}{2}(E+E^{\dag})$ is used in
the theory of Cooper pair box with a very thin junction when the
junction energy $E_J \cos \Phi$ is higher than the electrostatic
energy\cite{Bouchiat98}.

Further progress in the definition of phase operator was obtained
by Pegg and Barnett. The phase states are now defined from the
discrete Fourier transform (or more precisely the quantum Fourier
transform since the superposition is on Fock states not on real
numbers)
\begin{equation}
|\theta_p\rangle=\frac{1}{\sqrt{q}}\sum_{n=0}^{q-1}\exp(2i\pi\frac{p}{q}n)|n\rangle.
\label{QFT}
\end{equation}
The states are eigenstates of the Hermitian phase operator
\begin{equation}
\Theta_q=\sum_{p=0}^{q-1}\theta_p |\theta_p\rangle\langle
\theta_p|,\label{Pegg} \end{equation}
with $\theta_p =\theta_0+2\pi p/q$ and $\theta_0$ is a reference
angle. It is implicit in the definition (\ref{QFT}) that the
Hilbert space is of finite dimension $q$. The states
$|\theta_p\rangle$ form an orthonormal set and in addition the
projector over the subspace of phase states is
$\sum_{p=0}^{q-1}|\theta_p\rangle\langle \theta_p|=1_q$, where
$1_q$ is the unitary operator. Given a state $|F\rangle$ one can
write a probability distribution $|\langle \theta_p|F\rangle|^2$
which may be used to compute various moments, e.g. expectation
values, variances. The key element of the formalism is that first
the calculations are done in the subspace of dimension $q$, then
the limit $q \rightarrow \infty$ is taken\cite{Pegg89}.

We are now in position to define a quantum phase-locking operator.
Our viewpoint has much to share with the classical phase-locking
problem as soon as one reinterpret the fraction $\frac{p}{q}$ in
(\ref{QFT}) as arising from the resonant interaction between two
oscillators and the dimension $q$ as a number which defines the
resolution of the experiment. From now we emphasize such phase
states $|\theta'_p\rangle$ which satisfy phase-locking properties
and we impose the coprimality condition
\begin{equation}
(p,q)=1. \label{coprime}
\end{equation}
The quantum phase-locking operator is defined as
\begin{equation}
\Theta_q^{\rm{lock}}=\sum_p\theta_p |\theta'_p\rangle\langle
\theta'_p|,\label{Qlock}
\end{equation}
with $\theta_p=2\pi\frac{p}{q}$ and the notation $p$ means
summation from $0$ to $q-1$ with $(p,q)=1$. Using (\ref{QFT}) and
(\ref{coprime}) in (\ref{Qlock}) one obtains
\begin{equation}
\Theta_q^{\rm{lock}}=
\frac{1}{q}\sum_{n,l}c_q(n-l)|n\rangle\langle l|,
\label{projector}
\end{equation}
where the range of values of $n,l$ is from $0$ to $\phi(q)$, and
$\phi(q)$ is the Euler totient function. The coefficients in front
of the outer products $|n\rangle\langle l|$ are the so-called
Ramanujan sums
\begin{equation}
c_q(n)=\sum_p \exp(2i\pi \frac{p}{q}
n)=\frac{\mu(q_1)\phi(q)}{\phi(q_1)},~~\rm{with}~q_1=q/(q,n).
\label{Rsums}
\end{equation}
In the above equation $\mu(q)$ is the M\"obius function, which is
$0$ if the prime number decomposition of $q$ contains a square,
$1$ if $q=1$ and $(-1)^k$ if $q$ is the product of $k$ distinct
primes. Ramanujan sums are relative integers which are
quasi-periodic versus $n$ with quasi-period $\phi(q)$, and
aperiodic versus $q$ with a type of variability imposed by the
M\"obius function. Ramanujan sums have been used for signal
processing of low frequency noise\cite{PRE02}. With the Ramanujan
sum expansion the modified Mangoldt function introduced at the end
of Sect. \ref{Classical} is the dual of M\"obius function
\begin{equation}
b(n)=\frac{\phi(n)}{n}\Lambda(n)=\sum_{q\ge
1}\frac{\mu(q)}{\phi(q)}c_q(n) \label{dual}
\end{equation}
This illustrates that many \lq\lq interesting" arithmetical
functions carry the structure of prime numbers. We already
mentioned the relation $d \ln\zeta(s)/ds=\sum_{n\ge
1}\frac{\Lambda(n)}{n^s}$, but there is also the relation
$1/\zeta(s)=\sum_{n\ge 1}\frac{\mu(n)}{n^s}$. There is a well
known formulation of Riemann hypothesis from the summatory
M\"obius function $\sum_{n=1}^t \mu(n)=O(t^{1/2+\epsilon})$,
whatever $\epsilon$\cite{FNL01}.

Given a state $\beta$ one can calculate the expectation value of
the quantum phase-locking operator as
\begin{equation}
\langle \Theta_q^{\rm{lock}}\rangle=\sum_p \theta_p\langle
\theta'_p|\beta \rangle^2. \label{expec}
\end{equation}
If one uses the finite form of Susskind-Glogower phase states
(\ref{Susskind}) and a real parameter $\beta$
\begin{equation}
|\beta\rangle =\frac{1}{\sqrt{q}}\sum_{n=0}^{q-1}\exp(i n
\beta)|n\rangle,
\end{equation}
\begin{figure}[htbp]
\centering{\resizebox{7cm}{!} {\includegraphics{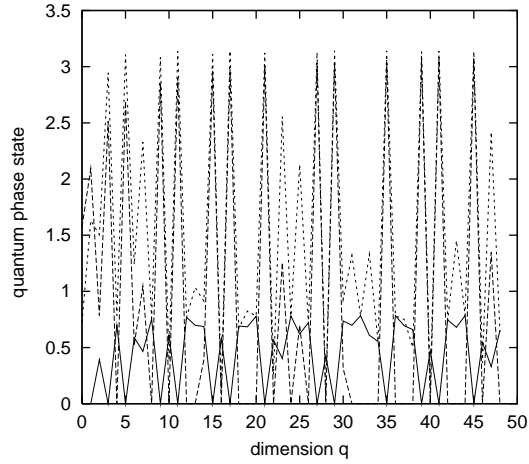}}}
\caption{Oscillations in the expectation value (\ref{expec2}) of
the locked phase at $\beta=1$ (dotted lines) and their squeezing
at $\beta=0$ (plain lines). The brokenhearted line which touches
the horizontal axis is $\pi \Lambda(q)/\ln q$}
\end{figure}
the expectation value of the locked phase becomes
\begin{equation}
\langle
\Theta_q^{\rm{lock}}\rangle=\frac{\pi}{q^2}\sum_{n,l}c_q(l-n)\exp(i\beta(n-l)).
\label{expec2}
\end{equation}
For $\beta=1$ it is found that $\langle
\Theta_q^{\rm{lock}}\rangle$ has the more pronounced peaks are at
such values of $q$ which are powers of a prime number. It can be
approximated by the normalized Mangoldt function $\pi
\Lambda(q)/\ln q$ as shown on Fig. 3. For $\beta=0$ the
expectation value of $\langle \Theta_q^{\rm{lock}}\rangle$ is much
lower. The parameter $\beta$ can be used to minimize the phase
uncertainty well below the classical value\cite{PLA03}.

A remarkable model of the quantum phase-locking effect and its
relation to prime number theory has been constructed by Bost and
Connes\cite{Bost95}. Instead of an ad-hoc quantum phase operator
as (\cite{Pegg89}) or (\ref{projector}), it is based on the
formulation of a dynamical system and its associated quantum
statistics. The dynamical system is first defined by an
Hamiltonian operator $H_0$ with eigenvalues equal to the
logarithms of integers
\begin{equation}
H_0|n\rangle= \ln n |n\rangle.
\end{equation}
Using the relations
$\exp(-\tilde{\beta}H_0)|n\rangle=\exp(-\tilde{\beta}\ln
n)|n\rangle=n^{-\tilde{\beta}}|n\rangle$, it follows that the
partition function of the model at the inverse temperature
$\tilde{\beta}$ is
\begin{equation}
\rm{Trace}(\exp(-\tilde{\beta
H_0}))=\sum_{n=1}^{\infty}n^{-\tilde{\beta}}=\zeta(\tilde{\beta}),
\end{equation}
where $\zeta(\tilde{\beta})$ is the Riemann zeta function already
met in Sec. \ref{Hyperbolic} in the definition of the hyperbolic
scattering coefficient (\ref{scatter}). In quantum statistical
mechanics, given an observable Hermitian operator $M$ one has the
Hamiltonian evolution $\sigma_t(M)$ versus time $t$
\begin{equation}
\sigma_t(M)=e^{itH_0}Me^{-itH_0},
\end{equation}
and the expectation value of $M$ is the Gibbs state \\
$\rm{Gibbs}(M)=\rm{Trace}(M
\exp(-\tilde{\beta}H_0))/\rm{Trace}(\exp(-\tilde{\beta}H_0)$.
\begin{figure}[htbp]
\centering{\resizebox{8cm}{!} {\includegraphics{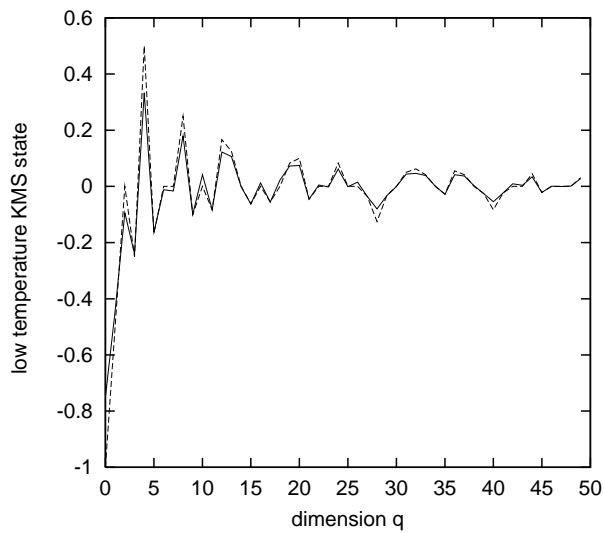}}}
\caption{Phase expectation value (\ref{KMS}) in Bost and Connes
model at the inverse temperature $\beta=3$ (plain lines) in
comparison to the function $\mu(q)/\phi(q)$ (dotted lines)}
\end{figure}
\begin{figure}[htbp]
\centering{\resizebox{8cm}{!} {\includegraphics{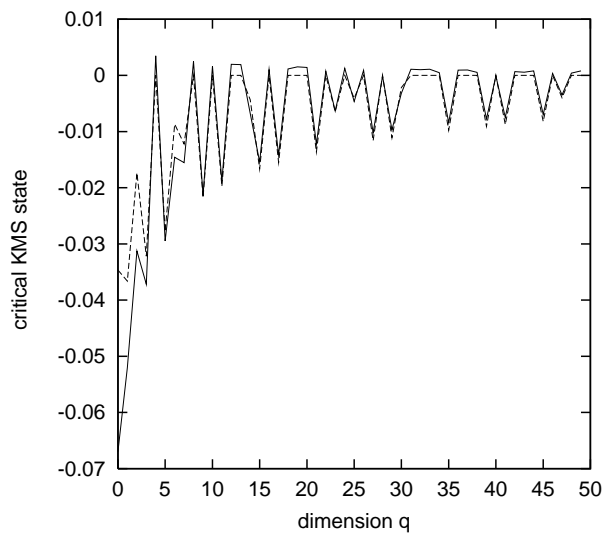}}}
\caption{Phase expectation value (\ref{KMS}) in Bost and Connes
model at the inverse temperature $\beta=1+\epsilon$,
$\epsilon=0.1$ (plain lines) in comparison to the function
$-\Lambda(q)\epsilon/q$ (dotted lines).}
\end{figure}
In Bost and Connes approach the observables belong to an algebra
of operators $\mu_q$ and $e_q^{(p)}$ which are defined by their
action on the occupation numbers $|n\rangle$ as
\begin{eqnarray}
&\mu_q|n\rangle =|qn\rangle, \label{shift}\\
&e_q^{(p)}|n\rangle=\exp(\frac{2i \pi p n}{q})|n\rangle.
\label{Qphase}
\end{eqnarray}
The first operator $\mu_q$ acts as a shift in the space of number
states; the second one is such that its action encodes the
individuals in the quantum Fourier transform (\ref{QFT}). Like in
the quantum phase-locking operator one uses the coprimality
condition (\ref{coprime}) to distinguish in (\ref{Qphase}) the
primitive roots of unity $\exp(2i \pi p/q)$, $(p,q)=1$. One can
show that there is a hidden symmetry group which is used to label
the elements of the algebra\footnote{This is the Galois group
$W=Gal(\mathcal{Q}^{\rm{cycl}}/\mathcal{Q})$ of the cyclotomic
extension on the field of rational numbers $\mathcal{Q}.$}. Using
the action of the group, the Gibbs state is replaced by the
so-called Kubo-Martin-Schwinger (or KMS) state. The system
exhibits a phase transition with spontaneous symmetry breaking at
the inverse temperature $\tilde{\beta}=1$ which corresponds to the
unique pole of the Riemann zeta function $\zeta(\tilde{\beta})$.
At low temperature $\tilde{\beta}>1$ one gets, after tricky
calculations, the expectation value of the phase operator which
replaces (\ref{expec2}) in the following form\cite{Bost95}
\begin{equation}
\rm{KMS}(e_q^{(p)})=q^{-\tilde{\beta}}\prod_{\stackrel{p~
\rm{divides}~ q }{p~
\rm{prime}}}\frac{1-p^{\tilde{\beta}-1}}{1-p^{-1}}. \label{KMS}
\end{equation}
The KMS state is represented for two limiting cases, the low
temperature limit $\beta\gg 1$ (Fig. 4) and the critical case
$\beta=1+\epsilon$ (Fig. 5), with $\epsilon\simeq 0$. In these
limits one has respectively $\rm{KMS}_{\beta\gg
1}(q)=\frac{\mu(q)}{\phi(q)}$ and $\rm{KMS}_{1+\epsilon}\simeq
\frac{-\Lambda(q)\epsilon}{q}$. In the low temperature limit the
spectrum (\ref{dual}) corresponding to the Ramanujan sum expansion
of the modified Mangoldt function $b(n)$ is recovered. Close to
the critical point $\beta=1+\epsilon$ the oscillations are
proportional to $\Lambda(q)\simeq b(q)$ and are of very small
amplitude due to the squeezing coefficient $\epsilon$. A
comparable squeezing effect was already observed in the
expectation value (\ref{expec}) of the quantum phase operator (see
Fig. 3). After the phenomenological model (\ref{flucgain}) and the
hyperbolic model (\ref{scatter}), the Bost and Connes cyclotomic
model also points to the Mangoldt function as a source of low
frequency fluctuations. In the last case the model is associated
to the spontaneous symmetry breaking and the squeezing of phase
oscillations at the critical KMS state.

As a conclusion phase-locking evokes Einstein \lq\lq spooky"
action at a distance. It is a very important concept in phase
sensitive communication circuits. Quantum phase-locking may become
synonymous of phase entanglement, be used to such tasks as remote
synchronization, quantum noise reduction and as a quantum resource
for the circuits of quantum communications.

\section*{Acknowledgements}
The author acknowledges Serge Perrine and Haret Rosu for their
collaboration and V.I. Man'ko for sending him his new book on
nonclassical states of light.

\end{document}